\documentclass[openacc]{rstransa}%%%%where rstrans is the template name

\usepackage[]{graphicx}

\titlehead{Research}

\begin{document}

%%%% Article title to be placed here
\title{Anisotropy in the cosmic acceleration inferred from supernovae}

\author{%%%% Author details
Mohamed Rameez}

%%%%%%%%% Insert author address here
\address{Tata Institute of Fundamental Research \\ Homi Bhabha Road, Colaba 400005\\Mumbai, Maharashtra, India
}

%%%% Subject entries to be placed here %%%%
\subject{Cosmology, Gravity, Fundamental Physics}

%%%% Keyword entries to be placed here %%%%
\keywords{Dark Energy, Cosmological Constant, Peculiar Velocities, Cosmological Principle, Bulk Flows}

%%%% Insert corresponding author and its email address}
\corres{Mohamed Rameez\\
\email{mohamed.rameez@tifr.res.in}}

%%%% Abstract text to be placed here %%%%%%%%%%%%
\begin{abstract}
Under the assumption that they are standard(isable) candles, the lightcurves of Type~Ia supernovae have been analyzed in the framework of the standard Friedmann-Lema\^itre-Robertson-Walker cosmology to conclude that the expansion rate of the Universe is accelerating due to dark energy. While the original claims in the late 1990's were made using overlapping samples of less than 100 supernovae in total, catalogues of nearly 2000 supernovae are now available. In light of recent developments such as the cosmic dipole anomaly and the larger than expected bulk flow in the local Universe (which does not converge to the Cosmic Rest Frame), we analyze the newer datasets using a Maximum Likelihood Estimator and find that the acceleration of the expansion rate of the Universe is unequivocally anisotropic. The associated debate in the literature highlights the artifices of using supernovae as standardisable candles, while also providing deeper insights into a consistent relativistic view of peculiar motions as departures from the Hubble expansion of the Universe. The effects of our being `tilted observers' embedded in a deep bulk flow may have been mistaken for cosmic acceleration. 
\end{abstract}
%%%%%%%%%%%%%%%%%%%%%%%%%%%

%%%%%%%%%% Insert the texts which can accomdate on firstpage in the tag "fmtext" %%%%%

\begin{fmtext}

%%%% Insert A head here

%\subsection{Insert B head here}
%%%% Insert B head here
%Subsection text here.

%\subsubsection{Insert C head here}
%%%% Insert C head here
%Subsubsection text here.

\end{fmtext}

%%%%%%%%%%%%%%% End of first page %%%%%%%%%%%%%%%%%%%%%

\maketitle

\section{Introduction}

Today's standard Lambda-Cold-Dark-Matter ($\Lambda$CDM) cosmological model was established over the past three decades through a process that was focused on getting diverse datasets to be concordant~\cite{Ostriker:1995su} when interpreted in the simplified framework of the Friedmann-Lema\^itre-Robertson-Walker (FLRW) cosmology. This assumed ~\cite{Weinberg:1972kfs} that the  Universe is sensibly isotropic and homogeneous on large scales: the Cosmological Principle  (CP). The dominant component of the energy density of this `concordance' model of the Universe is dark energy, represented in its most economical form by Einstein's Cosmological Constant $\Lambda$.

Evidence for late time acceleration of the expansion rate of the Universe was claimed in the late-1990's from observations of Type~Ia supernovae (SNe~Ia). High redshift SNe~Ia were found to be ``0.15 mag (15\% in flux) fainter than the low redshift supernovae'', compared to the expectation for a $\Lambda = 0$ Universe, in a sample of 42 high-z SNe~Ia from the Supernova Cosmology Project, fitted jointly with a sample of 18 SNe~Ia from the Calan/Tololo Supernova Survey~\cite{SupernovaCosmologyProject:1998vns}. Similar conclusions were drawn from an overlapping sample of 50 SNe~Ia (16 high-$z$ and 34 nearby)~\cite{SupernovaSearchTeam:1998fmf}. These findings, rewarded subsequently with the Nobel prize in physics (2011) ``for the discovery of cosmic acceleration'', were crucial in establishing $\Lambda$CDM as the standard cosmological model.

Recent developments warrant a reexamination of these conclusions. While the total number of SNe~Ia considered in the above studies was less than 100, many more have been observed over the past quarter  century, and the latest compilations have nearly $2000$ SNe~Ia~\cite{Rubin:2023ovl}. Moreover, it has been noted  that the ``constrained $\chi^2$ statistic'' employed in the original claims is unsuitable~\cite{Karpenka:2014uae, March:2011xa, Nielsen:2015pga, Mohayaee:2021jzi}. This has happened even as the `concordance' argument has begun to fall apart~\cite{DiValentino:2020hov}, with increasing tension between the present day value of the Hubble constant as inferred from probes of the early and late Universe~\cite{Riess:2019qba} --- leading to a so-called ``crisis in cosmology''.

Perhaps the most compelling reason to reexamine these claims is the  emerging consensus against the CP that underlies the FLRW cosmology. In this framework the observed dipole anisotropy of the Cosmic Microwave Background (CMB), which is $\sim 100$ times larger than the primordial CMB anisotropies observed at higher multipoles, is interpreted as a kinematic effect due to our local peculiar (i.e. non-Hubble) motion. By performing a special relativistic boost we should then be able to transform observational data to the Cosmic Rest Frame (CRF) in which the CMB should look isotropic. In the FLRW model the matter distribution should also be isotropic in this frame. The sky distribution of cosmologically distant sources as observed in our (heliocentric) frame should then also exhibit a similar dipole anisotropy \cite{1984MNRAS.206..377E}. This consistency test performed with multiple, independent ground and space based observatories at both radio~\cite{Wagenveld:2023kvi} and infrared~\cite{Secrest:2020has, Secrest:2022uvx} wavelengths has however revealed the `Cosmic dipole anomaly', namely a mismatch in the expected amplitude of the matter dipole. Moreover over the past two decades, the peculiar velocities in the local Universe have been mapped, using both direct distance measurements and indirect methods. As larger, deeper three-dimensional surveys of the local Universe became available, in addition to the uncorrelated velocities of the individual objects in the survey, evidence for a directionally coherent, large scale bulk flow has emerged. A reconstruction of the velocity field~\cite{Lavaux:2008th} from the three-dimensional 2MASS redshift survey data (2MRS) failed to find convergence to the CRF suggesting that matter out to $\sim150h^{-1}$~Mpc is still moving with a velocity $>100$~km\,s$^{-1}$. A similar bulk motion out to the Shapley supercluster was identified  using SNe~Ia as distance indicators~\cite{Colin:2010ds, Feindt:2013pma}. The direction of this bulk motion remains coherent between different shells. Such motions are indeed expected due to the growth of inhomogenieties as structure grows via gravitational instability around typical observers in the $\Lambda$CDM Universe. However the amplitude of the velocity field ought to die out as the size of the regions over which the peculiar velocities are averaged is increased. By contrast the observed peculiar flow of our part of the Universe appears to be too fast over too large a  distance. Analysis of the most up-to-date measurements of the distances and velocities of 38,000 groups and galaxies in the local Universe, the CosmicFlows 4 catalogue, concludes that our local Universe would be very rare in a gaussian random field, having less than 0.003\% probability of ocurring in a $\Lambda$CDM model~\cite{Watkins:2023rll}. 

If this bulk flow is real, it radically challenges the idea of cosmic acceleration due to dark energy, while also offering a simple explanation for the `Hubble tension'.

\subsection{Relativistic effects of the bulk flow}
When massive objects move, the flux of energy due to their motion in turn contributes to gravitational effects~\cite{Filippou:2020gnr}. This is a general relativistic effect with no Newtonian counterpart~\cite{Tsagas:2021dsl}, which has been excluded from standard cosmology on the grounds that any corrections should be small if the velocities are non-relativistic. However, studies of the impact of such effects on the rate of change of the expansion of space inferred by observers embedded within a  bulk flow~\cite{Tsagas:2015mua} suggest that even when the universe is globally decelerating, such `tilted observers' may infer accelerated expansion locally due to their drift motion, with the effect maximized in one direction of the sky and minimised in the opposite direction.
%(when the observers themselves are moving with respect to the bulk flow). 
A specific prediction is that the axis of this dipolar modulation of the cosmic deceleration parameter $q$ should lie fairly close to that of the CMB dipole axis and the dipole amplitude should decay with redshift~\cite{Tsagas:2011wq}. The same physical arguments have been exposited using both covariant and gauge invariant linear relativistic cosmological perturbation theory~\cite{Tsaprazi:2019bbi}, as well as the peculiar Raychaudhury equation~\cite{Tsagas:2013ila}, and generalized in the choice of background geometry~\cite{Tsagas:2021tqa}. Independent approaches have been suggested based on simulations employing numerical relativity~\cite{Heinesen:2021azp} and covariant cosmography~\cite{Maartens:2023tib} to study these effects, leading to similar conclusions about actual observables.

\subsection{Evidence for anisotropy of cosmic acceleration}

These predictions were tested~\cite{Colin:2019opb} using the SDSSII/SNLS3 Joint Lightcurve Analysis (JLA) Catalogue~\cite{SDSS:2014iwm} of data from 740 SNe~Ia. The local deceleration parameter $q_0$ was modified to include a scale-dependent dipolar component $q_0 = q_\mathrm{m} + \vec{q}_\mathrm{d} . \hat{n} {\rm exp}(-z/S)$.\footnote{It must be emphasised that this is a \emph{toy} model to test if the effects predicted for a tilted observer \cite{Tsagas:2011wq,Tsagas:2015mua} are present in the data. This model has the minimum number of additional parameters as is necessary to confront the limited data set.} Due to the uneven sky coverage of the JLA catalogue, the direction $\vec{q}_d$ was fixed to the CMB dipole direction in accordance with theoretical expectations. Since the formalism used to study the \emph{tilted Universe} pays particular attention to \emph{real observers}~\cite{Tsaprazi:2019bbi}, data as gathered in the heliocentric frame were employed in the analysis. When a maximum likelihood estimator \cite{Nielsen:2015pga} (which unlike the `constrained $\chi^2$' method usually employed in SNe~Ia data fitting, is suitable for model selection~\cite{Mohayaee:2021jzi}) was utilised to explore the parameter space, it was found that a large dipole anisotropy with $\vec{q}_\mathrm{d} \sim -8$ was preferred at $3.9 \sigma$ in frequentist statistical significance. The isotropic component of $q$ was found to have a value of $q_\mathrm{m} = -0.16$ (whereas in $\Lambda$CDM this should be -0.55), compatible with a non-accelerating universe at $1.4\sigma$. The best fit value of the scale parameter $S$ was found to be 0.0262 indicating that the anisotropic acceleration dominates over the isotropic component all the way out to $z \sim 0.1$. An \emph{a posteriori} test varying the direction of the dipole found the best-fit axis to be only $23^0$ away from the CMB dipole, suggesting strongly that what the data shows is in accordance with the expectation for the tilted Universe. General relativistic effects of peculiar velocities in the local Universe may have been mistaken for dark energy.

\section{The Dialectics of Nature}

These conclusions~\cite{Colin:2019opb} were challenged  immediately~\cite{Rubin:2019ywt}, but have been later reexamined independently by various authors~\cite{Rahman:2021mti, Dhawan:2022lze}. While a more detailed response can be found in Ref.~\cite{Mohayaee:2021jzi}, we outline here the two most important sources of dispute.

\subsection{Issue 1 : Sample and Redshift dependence of SNe~Ia standardisation}

Type Ia Supernovae are believed to be thermonuclear explosions in low-mass stars, e.g. triggered when the mass of a Carbon-Oxygen white dwarf is driven, by the accretion of material from a companion, over the maximum that can be supported by electron degeneracy pressure. Since this happens near a critical mass, the Chandrasekhar limit of $\sim 1.4 M_\odot$, all SNe~Ia are taken to have the same intrinsic luminosity, i.e. a `standard candle'. In practice the intrinsic magnitudes of nearby SNe~Ia (to which distances are known via independent means) exhibit a rather large scatter. However by exploiting the observed linear correlation of the (colour-dependent) luminosity decline rate with the peak magnitude~\cite{Phillips:1993ng}, this scatter can be considerably reduced. This makes SNe~Ia `standard(isable)' candles, i.e. the intrinsic magnitude can be inferred with relatively low scatter ($\sim0.1-0.2$ mag) by measuring the lightcurves in different (colour) bands~\cite{Leibundgut:2000xw}. Further assuming that the intrinsic properties themselves do not evolve with redshift, one may \emph{hope} to use observations of SNe~Ia to measure the cosmological evolution of the luminosity distance (i.e. of the scale factor) as a function of redshift.

Early discordance (see Ref.~\cite{Bengochea:2010it} and Figure~4 of Ref.~\cite{Leibundgut:2000xw}) between the different empirical techniques for implementing the Phillips corrections~\cite{Phillips:1993ng}, viz. the Multi Colour Lightcurve Shape (MLCS) strategy~\cite{SupernovaSearchTeam:1998fmf}, the `stretch factor' corrections~\cite{SupernovaCosmologyProject:1998vns} and the template fitting or $\Delta m_{15}$ method~\cite{Hamuy:1996ss,Phillips:1999vh} has given way to the `Spectral Adaptive Lightcurve Template' (SALT), a two-step process wherein the shape as well as the colour \cite{Tripp:1997wt} parameters required for the standardisation are first derived from the lightcurve data, and the cosmological parameters are then extracted in a separate step \cite{Guy:2005me}. The current incarnation of this method is SALT2, employed in analysis of recent SNe~Ia data sets~\cite{SDSS:2014iwm, Pan-STARRS1:2017jku,Scolnic:2021amr}, in which every SNe~Ia is assigned three parameters, $m_B^*$, $x_1$ and $c$ --- respectively the apparent magnitude at maximum (in the rest frame `B-band'), the lightcurve shape, and the lightcurve colour correction. This can be used to construct the distance modulus using the Phillips-Tripp formula \cite{Tripp:1997wt}:

%In detail however the different empirical techniques for implementing the Phillips corrections~\cite{Phillips:1993ng}, viz. the Multi Colour Lightcurve Shape (MLCS) strategy~\cite{SupernovaSearchTeam:1998fmf}, the `stretch factor' corrections~\cite{SupernovaCosmologyProject:1998vns} and the template fitting or $\Delta m_{15}$ method~\cite{Hamuy:1996ss,Phillips:1999vh}, do \emph{not} agree with each other --- see Figure~4 of Ref.~\cite{Leibundgut:2000xw}. As the numbers of SNe~Ia has grown, the tension between the methods has in fact  increased~\cite{Bengochea:2010it}. The MLCS strategy was to simultaneously infer the Phillips corrections and the cosmological parameters using Bayesian inference. However a two-step process, the `Spectral Adaptive Lightcurve Template' (SALT), is now adopted, wherein the shape as well as the colour \cite{Tripp:1997wt} parameters required for the Phillips corrections are first derived from the lightcurve data, and the cosmological parameters are then extracted in a separate step \cite{Guy:2005me}. The current incarnation of this method is SALT2, employed in analysis of recent SNe~Ia data sets~\cite{SDSS:2014iwm, Pan-STARRS1:2017jku}, in which every SNe~Ia is assigned three parameters, $m_B^*$, $x_1$ and $c$ --- respectively the apparent magnitude at maximum (in the rest frame `B-band'), the lightcurve shape, and the lightcurve colour correction. This can be used to construct the distance modulus using the Tripp formula \cite{Tripp:1997wt}:

\begin{equation}
\label{eq:distmod}   
\mu_\mathrm{SN} = m_B^* - M^0_B + \alpha x_1 - \beta c ,
\end{equation}
\noindent
In our analysis~\cite{Colin:2019opb}, we employed distance modulii constructed uniformly as above for the entire dataset, following the analysis~\cite{SDSS:2014iwm} with which the data were originally disseminated. However, Ref.~\cite{Rubin:2019ywt}, following Ref.~\cite{Rubin:2016iqe}, argue that the distributions of $x_1$ and $c$ must be parameterized in such a way as to include supernova sample as well as redshift dependence, leading to a proliferation (doubling) in the number of total parameters. 
%While a subsample dependence to these distributions (though originally not included in Ref.~\cite{SDSS:2014iwm}), may be justified, 

Allowing the distributions of $x_1$ and $c$ to be redshift dependent diminishes the case for SNe~Ia being standard(isable) candles for cosmology. This can be appreciated from the fact that when tracing out the expansion history of the Universe using SNe~Ia, one is hoping that $\text{d}\mu_\mathrm{SN}/\text{d}z$ is a faithful proxy for $\text{d}\mu_\mathrm{th}/\text{d}z$, where $\mu_\mathrm{th}$ is the theoretical distance modulus in an FLRW Universe. By allowing $x_1$ and $c$ to be redshift dependent, we allow $\text{d}\mu_\mathrm{SN}/\text{d}z$  to be further modified additively by $\alpha(\text{d}x_1/\text{d}z)-\beta(\text{d}c/\text{d}z)$. Apart from the contrived possibility of an exact cancellation (wherein $\alpha(\text{d}x_1/\text{d}z)=\beta(\text{d}c/\text{d}z)$), this merely allows for an additive modification of $\text{d}\mu_\mathrm{SN}/\text{d}z$. This is moreover done \emph{a posteriori} and in contravention of the choices of the very analysis with which the dataset was disseminated~\cite{SDSS:2014iwm}; also there is no evidence to support the convenient conspiracy that $\alpha(\text{d}x_1/\text{d}z)=\beta(\text{d}c/\text{d}z)$.

If the intrinsic properties of SNe~Ia (such as $M^0_B$) were to evolve with redshift this would of course trivially undermine the inference of accelerated expansion \cite{Tutusaus:2017ibk,Tutusaus:2018ulu}. In fact there are now independent indications~\cite{Kang:2019azh, Lee:2020usn} that SNe~Ia may not be sufficiently standardisable to faithfully map out the expansion history of the Universe. However there has been no suggestion of any dependence of the intrinsic properties of SNe Ia on direction, hence \emph{prima facie} one can still use them for studying anisotropies of the expansion rate.

\subsection{Issue 2 : Peculiar velocity 'corrections' and a shell crossing singularity}

Our analysis~\cite{Colin:2019opb} employed heliocentric observables in the fit, since predictions regarding the general relativistic effects of the local bulk flow have been made in an observer-dependent manner, and the heliocentric frame is the closest to a single observer frame for the whole dataset. Ref.~\cite{Rubin:2019ywt} described this choice as ``shocking'', as it allows the ``well-established motion of the solar system with respect to the CMB to imprint on the SN redshifts''. 

The redshift corrections for peculiar velocities are carried out according to the formula \cite{Ellis:1987zz}
\begin{equation}
\label{eq:zpv}   
1+z = (1 + z_\mathrm{O})(1 + z_\mathrm{c})(1 + z_\mathrm{s})
\end{equation}
\noindent
Here, the 'cosmological redshift' ($z_\mathrm{c}$, attributed to the scale factor evolution in the CRF of the background FLRW cosmology) is modified by Doppler red/blue shifting due to the peculiar velocity of both the source ($z_\mathrm{s}$), and of the observer ($z_\mathrm{O}$), with respect to the CRF.

These specific corrections, in particular for the motions of the sources with respect to the CRF were first adopted in SNe~Ia cosmology~\cite{SNLS:2011lii} only in 2011, following Ref.~\cite{Davis:2010jq}, which noted that until then ``less accurate'' corrections (based on just adding the redshifts) had been employed, and that too just for the motion of the observer with respect to the CRF.\footnote{Back in 1987, Ref.~\cite{Ellis:1987zz} had already discussed these corrections more generally, as part of the process of choosing the `corresponding 2-spheres' when fitting cosmological data.}

In Figure 2 of Ref.~\cite{Colin:2019opb} and the associated discussion, we have explained why we prefer to \emph{not} use these corrections as they ship with the disseminated data. While the interpretation of the CMB dipole as being purely kinematic because of our peculiar velocity of 369 km\,s$^{-1}$ wrt to the CRF (which goes into computing $z_\mathrm{O}$ for each supernova) stands challenged by the cosmic dipole anomaly) \cite{Secrest:2020has,Secrest:2022uvx,Wagenveld:2023kvi}, the lack of convergence to the CRF in the bulk flow of the local Universe \cite{Lavaux:2008th,Watkins:2023rll} makes the corrections with $z_\mathrm{s}$ poorly conceived. In particular, the model employed for such corrections~\cite{Hudson:2004et} in the JLA dataset reports a residual bulk flow of $687\pm203$~km\,s$^{-1}$ for the whole survey volume. As we illustrated in Figure 2 of Ref.~\cite{Colin:2019opb}, while SNe~Ia which are inside the survey volume were corrected for the motions of their host galaxies w.r.t. the CRF, the ones immediately outside were left uncorrected, introducing an arbitrary discontinuity within the data. 
Quite literally, the data as publicly disseminated encoded a picture in which a local spherical volume of 120$h^{-1}$ Mpc is smashing, at $687\pm203$~km\,s$^{-1}$ into the rest of the universe which is arbitrarily treated as at rest (with respect to the CRF).

If such a thing happened in one of the relativistic numerical/N-body/hydrodynamic simulations of the Universe, it would be called a shell crossing singularity. It is a sign of the system having evolved beyond the regime of validity of the numerical approximations being employed in the simulation (thus leading to unphysical outcomes). This is something to be avoided, but in the dataset which forms the primary empirical support for what has been called ``arguably the most important problem in theoretical physics'', this had been inserted by hand, and as we demonstrated in Figures 3 and 4 of Ref.~\cite{Mohayaee:2021jzi}, this arbitrary discontinuity constitutes about half the evidence for cosmic acceleration.~\footnote{This can be appreciated from the fact that the shift in magnitude associated with the discontinuity $\Delta\mu = 5v/\mathrm{log}(10)cz$ is $\sim0.07$ mag for the $687\pm203$~km\,s$^{-1}$ bulk flow extending out to $120h^{-1}$ Mpc in the SMAC~\cite{Hudson:2004et} sample.} Consequently, as can be seen in these figures, as the two parts of the peculiar velocity corrections are introduced, the best-fit value of $q_0$ moves from a value that is compatible with 0, towards more negative values indicative of cosmic acceleration.

\section{The sculpting of an elephant}

\begin{quote}
\textit{`What is a fool-proof method for sculpting an elephant? First you get a block of granite; then you chip away everything that does not look like an elephant'}
\end{quote} 

\begin{figure}[!h]
\centering\includegraphics[width=3.4in]{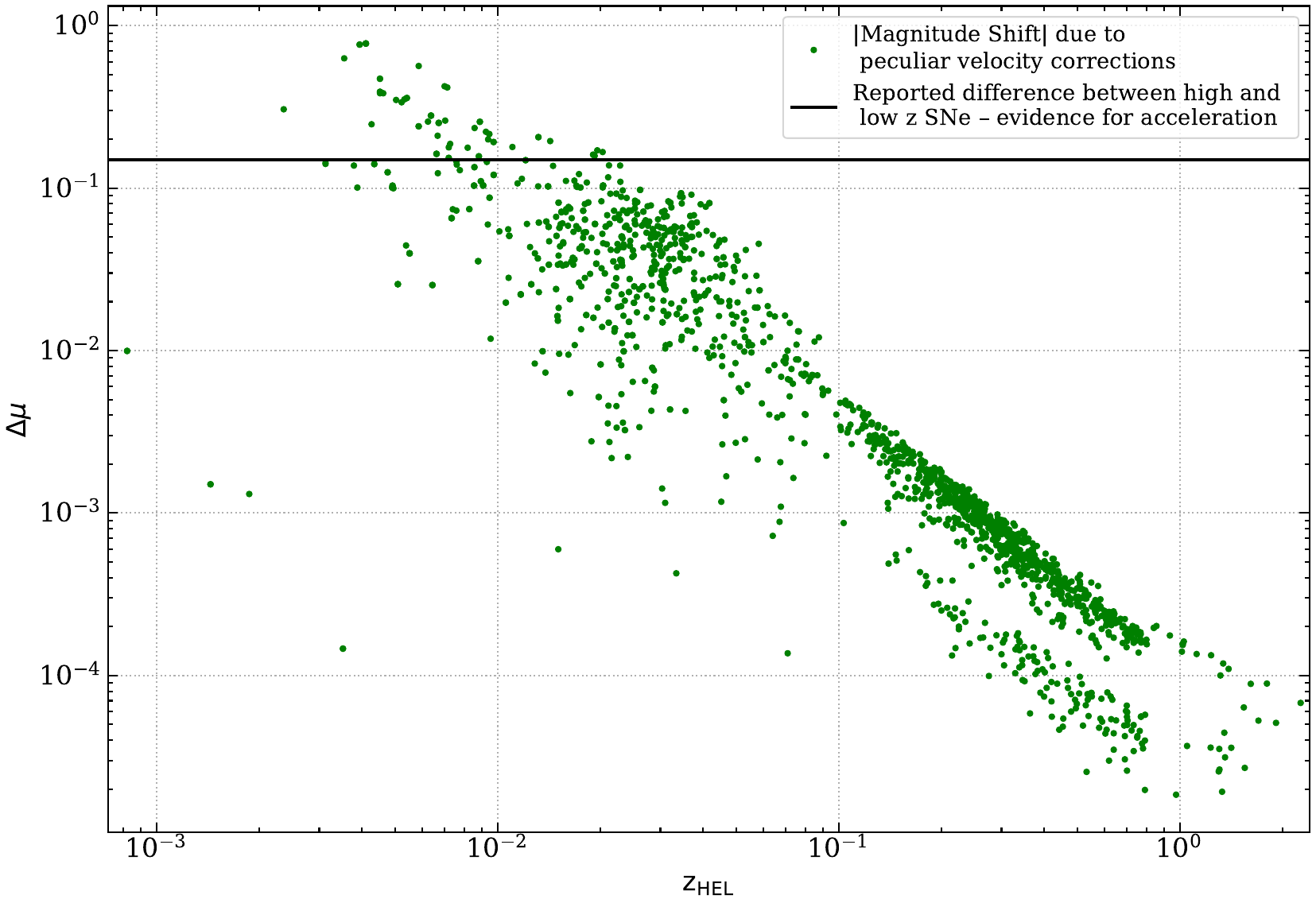}
%%% where xxxxxx name represents ''figurename.eps''
\caption{The shift in the magnitudes of Pantheon+ SNe~Ia caused by the applied peculiar velocity corrections ($\Delta\mu = {5v}/{\mathrm{log}(10)cz}$) versus redshift. The signal for cosmic acceleration, viz. the 0.15 mag dimming of high $z$ SNe~Ia w.r.t. the low $z$ ones, is shown for comparison as a black horizontal line.}
\label{fig_ele}
\end{figure}

This is now acknowledged to have been a mistake. The subsequent Pantheon+ compilation~\cite{Carr:2021lcj} (see section 6.3), remarks that the ``peculiar velocities of galaxies outside $r_\textrm{max}$ should not be set to zero''. In order to ``ensure a smooth transition across $r_\mathrm{max}$'', they now ``choose to model the bulk flow as a decaying function consistent with $\Lambda$CDM expectations. 
In Pantheon+, the redshifts of \emph{all} SNe~Ia have to be thus corrected, both for the motion of the observer w.r.t. the CRF, as well as that of the host galaxies of the SNe~Ia w.r.t. the CRF. 

It is worth emphasizing that peculiar velocities are \emph{defined} as the residual velocities of objects after their velocities due to the putative isotropic Hubble expansion have been subtracted out. In particular, the model employed in Pantheon+ to correct for peculiar velocities in the local Universe~\cite{Carrick:2015xza}, utilizes linear Newtonian perturbation theory to infer the peculiar velocity field from the 2M++ density contrast field, explicitly \emph{assuming} a $\Lambda$CDM universe. %\footnote{I first came across the joke within the quotation at the beginning of this subsection back in 1997, when I was 8 years old. However I could not understand it then. It is only after I became aware of how SNe~Ia data have to be corrected for peculiar velocities before they are fitted to data, did I understand the joke in its full potential.} 
 Note that this model also reports with a statistical significance $5.3\sigma$, a bulk flow of $159\pm23$~km\,s$^{-1}$ extending beyond the survey limit. The relevance of the quotation at the top of this subsection must now be clear.

\section{Anisotropy in the Pantheon+ compilation}

\begin{figure}[!h]
\centering\includegraphics[width=3.1in]{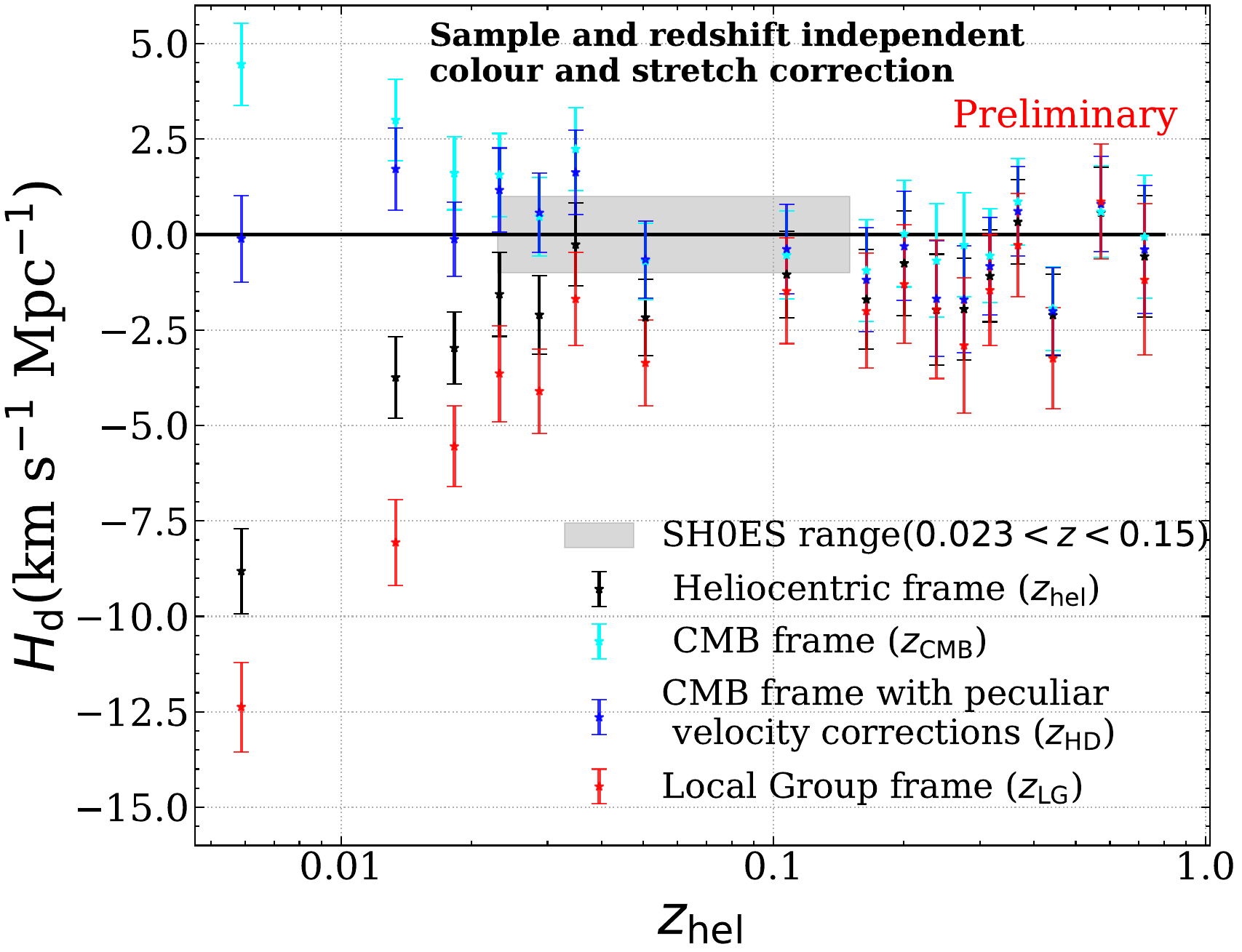}
\caption{Dipole in the Hubble expansion rate, extracted for 17 distinct redshift shells each containing 100 SNe~Ia from the Pantheon+ compilation. The redshift range in which the SH0ES measurements are performed is shaded in gray, with its vertical spread indicating the claimed uncertainty on $H_0$. Different colours correspond to redshift corrections for different choices of observer frames and peculiar velocities.
%While these results are obtained allowing sample and redshift dependence of the lightcurve stretch $x_1$ and colour $c$ corrections, the answer is similar even when these parameters are held constant. 
See Ref.~\cite{Sah:2024csa} from which this figure is taken, for more details.}
\label{fig_PHd}
\end{figure}

\begin{figure}[!h]
\centering\includegraphics[width=3.0in]{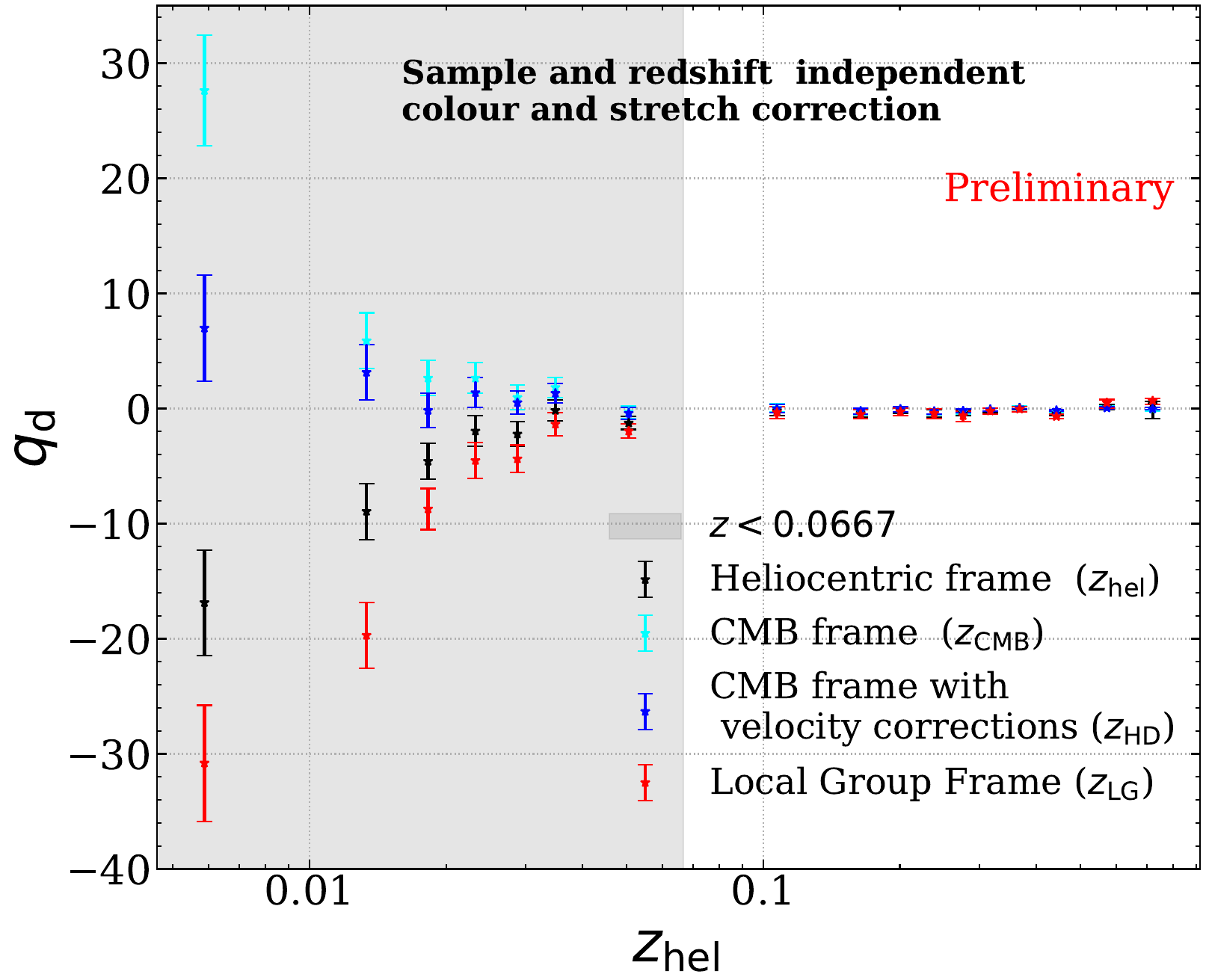}
\caption{Dipole in the deceleration parameter, extracted for 17 distinct redshift shells each containing 100 SNe~Ia from the Pantheon+ compilation. 
%While these results are obtained allowing sample and redshift dependence of the lightcurve stretch $x_1$ and colour $c$ corrections, similar conclusions are obtained even when these parameters are held constant. 
See Ref.~\cite{Sah:2024csa} from which this figure is taken, for more details.}
\label{fig_PQd}
\end{figure}

The Pantheon+ compilation includes over 400 new SNe~Ia at $z <0.1$ in comparison to JLA, allowing studies such as in Ref.~\cite{Colin:2019opb} to be carried out tomographically in redshift shells \cite{Sah:2024csa}; see also Ref.~\cite{McConville:2023xav}. The results of fits for a scale-independent dipole in the Hubble parameter are illustrated in Fig.~\ref{fig_PHd}, while similar fits for a dipolar modulation in the deceleration parameter are shown in Fig.~\ref{fig_PQd}.

The anisotropy seen in $q_0$ in the low redshift shells confirms the findings of Ref.~\cite{Colin:2019opb}, reiterating that the cosmic acceleration seen in SNe~Ia data is a frame-dependent, anisotropic effect. The anisotropy is strongest in the Local Group frame, and vanishes only if the redshifts are corrected also for the peculiar velocities of the host galaxies of the SNe~Ia. The correction for the motion of the heliocentric frame with respect to the CRF merely flips the sign of the dipole (since the motion of the Sun around the Galaxy is in nearly the opposite direction to the CMB dipole). 

It is particularly noteworthy that in the redshift shells towards the lower end of the $0.023 < z< 0.15$ range in which SH0ES measures the local value of the Hubble parameter, the best fit values of the dipole in the Hubble parameter are larger than the uncertainty on $H_0$ quoted by SH0ES \cite{Riess:2019qba}.

\section{Peculiar velocities, the fitting problem in Cosmology and the myth of the Cosmic Rest Frame}

In Refs.~\cite{Davis:2010jq, SNLS:2011lii, Colin:2010ds, Carrick:2015xza, Watkins:2023rll} peculiar velocities are considered to be motions of matter with respect to an isotropically expanding space associated with the background FLRW cosmology. However, since General Relativity is a \emph{background independent} theory, it is more correct to think of them as differences in the expansion velocity field of the Universe itself~\cite{McClure:2007vv}, i.e. a differential expansion of space in an everywhere different Universe. In this latter view, the peculiar velocity `corrections' applied to SNe~Ia data take on a more sinister connotation. They are merely a way to correct for the `true shape' of the expansion velocity field of the local Universe and thus isotropise the data, in the process inserting the relative 0.15 mag dimming of high-$z$ SNe~Ia into the dataset (see Fig.\ref{fig_ele}), before it is fitted to the isotropic-homogeneous FLRW cosmological model. 

What the bulk flow observed in the local Universe actually indicates is that \emph{the real Universe is anisotropic out to at least 200$h^{-1}$ Mpc}, approximately a billion light years. The `Great Attractor' has however not yet been found, and there is no convergence yet to the Cosmic Rest Frame.
%must then be understood as a myth, much like the idea of a center to the Universe in pre-Copernican times. 
It is the only meaningful frame in an FLRW spacetime, yet we seem unable to find it in the real Universe. These observations take on deeper significance when considered within the context of the debate concerning the effect of inhomogeneities in cosmology~\cite{Ellis:2005uz}.

In discussing `the fitting problem in cosmology', Ref.~\cite{Ellis:1987zz} presented the perspective that fitting an FLRW model to data from the real Universe is like fitting a perfect sphere to the  Earth. The validity of the latter representation is clearly a question of precision. While the Earth can meaningfully be considered a sphere to a precision of ${\cal O}(50)$~km on the radius, if further precision (${\cal O}(5)$~km) is pursued then the sphere description breaks down and a more detailed picture of an oblate spheroid with mountains, valleys and craters emerges. This analogy provides a simple framework for understanding the last few decades in cosmology and the ongoing crisis due to the Hubble tension. The FLRW model was fine as long as the Hubble parameter could be measured only to $\sim10\%$ precision. However the real Universe is clearly not FLRW to a precision of 1\% on $H_0$, as can be seen in Fig.~\ref{fig_PHd}. It is then natural to predict that in the near future, as more data becomes available, $H_0$ will be established to vary systematically across the sky by more than the small uncertainty claimed by SH0ES, as long as the data are not ``corrected'' \emph{post facto} for peculiar velocities.

The anisotropy of the local Universe is thus an enduring feature stretching far out into the cosmological sky, unchanging over timescales relevant to human activity,\footnote{This can be appreciated by considering the rate at which peculiar (and Hubble) velocities are expected to change in reality, from either linear theory~\cite{1980lssu.book.....P}, or empirical constraints on the redshift drift~\cite{Bolejko:2019vni}.} and the peculiar velocity corrections applied to SNe~Ia data are thus, a way of ``untilting the Universe''.

\section{Discussion and Conclusions}

In this brief review we have attempted to provide a critical perspective on the inference of dark energy from SNe~Ia data. We have provided an alternative  explanation~\cite{Colin:2019opb,Sah:2024csa} using known (but not widely appreciated) physics, in the form of General Relativistic effects of the local bulk flow. We have demonstrated that some of the specific issues introduced in the subsequent debate undermine the basic assumption that SNe~Ia are sufficiently standardisable candles as to be useful for precision cosmology. 

\begin{table}
\centering
\resizebox{\textwidth}{!}{%
\begin{tabular}{|l|c|c|c|c|c|c|}
\hline
Collaboration &Number of SNe~Ia &$N_\mathrm{out}$ &Lightcurve &CRF &Treatment of peculiar velocities &Lensing\\
\hline
SCP \cite{SupernovaCosmologyProject:1998vns} &60 (18+42) &4 &``stretch'' &LG &$\sigma_v = 300$~km~s$^{-1}$ &\\
HZT \cite{SupernovaSearchTeam:1998fmf} &50 (34+16) &- &MLCS, template &CMB &$\sigma_v = 200$~km~s$^{-1}$, 2500~km~s$^{-1}$ at high $z$&\\
SNLS \cite{SNLS:2005qlf} &117 (44+73) &2$^a$ &SALT &CMB+helio &$z_\mathrm{min}=0.015$&\\
SCP (Union) \cite{SupernovaCosmologyProject:2008ojh} &307&8$^a$ &SALT &CMB+helio &$z_\mathrm{min}=0.015$, $\sigma_v = 300$~km~s$^{-1}$  &$\sigma^l=0.093z$\\
Union2 \cite{Amanullah:2010vv} &557 &12$^a$ &SALT2 &CMB+helio &$z_\mathrm{min}=0.015$, $\sigma_v = 300$~km~s$^{-1}$ &$\sigma^l=0.093z$\\
SCP \cite{SupernovaCosmologyProject:2011ycw} &580 &0 &SALT2 &CMB+helio &not available &corrections\\
SNLS \cite{SNLS:2011lii} &472 &6$^a$ &SALT2 \& SiFTO &CMB &$\sigma_v = 150$~km~s$^{-1}$ + SN-by-SN corrections &$\sigma^l=0.055z$\\
JLA \cite{SDSS:2014iwm} &740 &0 &SALT2 &CMB &$\sigma_v = 150$~km~s$^{-1}$ + SN-by-SN corrections &$\sigma^l=0.055z$\\
Pantheon$^b$ \cite{Pan-STARRS1:2017jku} &1048 &86 &SALT2 &CMB &$\sigma_v = 250$~km~s$^{-1}$ + SN-by-SN corrections &$\sigma^l=0.055z$\\
Pantheon+ \cite{Scolnic:2021amr} &1701 (1550 unique) &433 &SALT2 &CMB &$\sigma_v = 250$~km~s$^{-1}$ + SN-by-SN corrections + analytic form beyond $r_{max}$ &$\sigma^l=0.055z$\\
Union3$^c$ \cite{Rubin:2023ovl} &2087&$624$ &SALT3 &CMB & Theoretical covariances\protect\footnotemark &$\sigma^l=0.055z$\\
\hline
\end{tabular}
}
\caption{Summary of SNe~Ia cosmology analyses. $N_\mathrm{out}$ is the number of outliers rejected, while CRF is the choice made of Cosmic Rest Frame. The first analyses argued that gravitational lensing is negligible, however subsequently a $z$-dependent systematic uncertainty has been introduced to account for it. This is an updated version of Table`1 in Ref.~\cite{Mohayaee:2021jzi}.\\
$^a$ Explicitly noted as ``3 $\sigma$ outlier'' rejections.\\
$^b$ While not explicitly documented, the Pantheon compilation initially had `corrections' for peculiar velocities applied far beyond the extent of the model \cite{Rameez:2019wdt}
%presumably just as in the subsequent Pantheon+ compilation. 
While this issue has been fixed on \href{https://github.com/dscolnic/Pantheon/issues/}{github}, it has not been documented in the literature. See Ref.~\cite{Rameez:2019wdt} for a discussion.\\
$^c$ This dataset has not been made public in a manner in which it can be independently reanalyzed to assess  the impact of peculiar velocities.}
\label{tab:SNeIasummary}
\end{table}

Of particular interest in the context of recent challenges to the Cosmological Principle is the question of how SNe~Ia data are corrected for peculiar velocities in the local Universe, which include a directional bulk component which has not been found to converge to the CRF. We have argued that these ``corrections'' artificially introduce the signal for acceleration into data
%through circular reasoning around a pre-assumed $\Lambda$CDM cosmology, 
and that the lack of convergence of the local bulk flow to the CRF in the very models used to correct SNe~Ia data indicate a fundamental inconsistency. These claims must be viewed within the context of how choices in SNe~Ia data analyses have evolved over the past quarter of a century while the number of SNe~Ia were building up from less than 100 to almost 2000 (see Table ~\ref{tab:SNeIasummary}). While we find (see Figs. 3, 5 \& 6 of Ref.~\cite{Sah:2024csa}) much like other authors~\cite{Horstmann:2021jjg, Sorrenti:2022zat} who have examined SNe Ia data, that the higher redshift supernovae show no significant anisotropy and thus agree with the general picture of $\Lambda$CDM cosmology, it is the \emph{comparison} in brightness of the high-z SNe with the low-z ones which provide the crucial evidence for cosmic acceleration due to $\Lambda$.

Such considerations must be viewed in conjunction with the recent challenge to the CP in the form of the cosmic dipole anomaly~\cite{Secrest:2020has, Secrest:2022uvx, Wagenveld:2023kvi} which independently questions the idea of the CMB frame as the CRF. In this context it has been argued that the differential expansion of space produced by nearby nonlinear structures cannot be reduced to a local boost with respect to the rest frame of an FLRW cosmology~\cite{Wiltshire:2012uh}.

Our view is that while the CP was a reasonable assumption in the late 1990s (because no data existed then to suggest otherwise), it should be abandoned now in favour of first carrying out detailed modelling of the local Universe over the introduction of exotic components of the energy density assuming an overly simple model universe. That the Pantheon+ analysis~\cite{Scolnic:2021amr} chooses instead to correct \emph{all} SNe~Ia for a directional bulk flow before fitting to an assumed FLRW cosmology highlights the fundamental inconsistency in the present approach.

\section{The future: Rubin-LSST and DESC} %Project 254 : ''Testing tilted cosmology''}
\footnotetext{In the recent analysis of the Union 3 catalogue~\cite{Rubin:2023ovl}, the treatment of peculiar velocities seems to rely on the covariances computed in Ref.~\cite{Hui:2005nm}. This has been done earlier to argue that bulk flows are a ``guaranteed theoretical signal'' which should be accounted for when testing cosmology~\cite{Huterer:2015gpa}. By including the additional covariances, the statistical preference for a bulk flow can be artificially weakened (see Figure 3 of Ref.~\cite{Huterer:2015gpa}) to then claim ``no evidence for bulk velocity \ldots''. However  these covariances are irrelevant for cosmological inference, as they quantify the spread in magnitude due to peculiar velocities as viewed by \emph{multiple typical} (i.e. Copernican) observers in the $\Lambda$CDM universe (see Figure 1 in Ref.~\cite{Mohayaee:2020wxf} and associated discussion), whereas the real Universe can be observed only from our unique vantage point. The correlations we expect between supernovae in a JLA-like catalogue in a local Universe like environment are 2–-8 times stronger than seen by a typical or Copernican observer \cite{Mohayaee:2020wxf}. This illustrates how peculiar velocity corrections can have a large impact on the value of the Cosmological Constant inferred from supernova data.}

Nevertheless the cosmology community still accepts $\Lambda$CDM as its standard model. The forthcoming Legacy Survey of Space and Time (LSST) at the Vera C. Rubin observatory provides an excellent opportunity to examine these important issues in an unprejudiced manner. Our simulations suggest that a sample of $\sim 5000$ SNe~Ia over the $\sim40\%$ of the sky to be surveyed will be sufficient to establish the existence of a scale-dependent dipolar modulation in the local deceleration parameter at $>5\sigma$ statistical significance. Since Rubin-LSST~\cite{LSSTScience:2009jmu} is expected to provide tens of thousands of SNe~Ia every year, this debate should be soon settled. 
%To this end we introduce Project 254, "Testing tilted cosmology" in the Dark Energy Science Collaboration. 
To do so, it is important however to produce an SNe~Ia data pipeline that is as free as possible of assumptions concerning the CRF, and then perform a principled, blinded analysis. 

\ack{I thank Animesh Sah for the figures, and the referees for insightful comments which helped significantly improve the draft. This work is the product of insights from many discussions over the last five years, mainly with Subir Sarkar, Roya Mohayaee, Jacques Colin, Sebastian von Hausegger, Nathan Secrest, Christos Tsagas, Konstantinos Migkas, Roy Maartens, Rocky Kolb, Arman Shafieloo, Harry Desmond, Jessica Santiago, Kerkyra Asvesta, Timothy Clifton, Pratyush Pranav, Jenny Wagner, David Wiltshire, Asta Heinesen, Lawrence Dam, Arman Shafieloo, Eoin O'Colgain \& Wendy Gray.}

\newpage
%%%%%%%%%% Insert bibliography here %%%%%%%%%%%%%%
\bibliographystyle{unsrt} \bibliography{sample}
%\begin{thebibliography}{9}
%\end{thebibliography}

\end{document}